\documentclass[
aps,
prd,
showpacs,
twocolumn,
superscriptaddress,
nofootinbib,
preprintnumbers
]{revtex4-1} 

\makeatletter
\def\p@subsection{}
\makeatother

\usepackage{color,graphicx,amsmath,amssymb,lipsum}
\usepackage[colorlinks=true,citecolor=blue,linkcolor=blue,urlcolor=blue, backref=false,pdfborder={0 0 0}]{hyperref}

\usepackage[normalem]{ulem}
\usepackage[utf8]{inputenc} 
\usepackage{mathtools}
\usepackage{tabularx} 
\usepackage{tabu}
\usepackage{multirow}
\usepackage[table]{xcolor}
\usepackage{colortbl}
\usepackage{caption}

\usepackage{float}

\usepackage{graphics,graphicx,tikz,tensor}
\usetikzlibrary{arrows,decorations.pathmorphing,patterns}

\usepackage{bm}

\newcommand{\be}{\begin{equation}}
\newcommand{\ee}{\end{equation}}
\newcommand{\tabeq}[2]{ \parbox{#1}{  \be\begin{aligned}#2 \end{aligned} \nonumber \ee }}

\begin{document}

\preprint{QMUL-PH-25-09}

\title{
Nonlinear Gravitational Memory in the Post-Minkowskian Expansion
}

\author{Alessandro Georgoudis}
\affiliation{Centre for Theoretical Physics, Department of Physics and Astronomy,\\
			Queen Mary University of London, 
			Mile End Road, London, E1 4NS, United Kingdom}
\author{Vasco Goncalves}
\affiliation{Centro de Fisica do Porto e Departamento de Fisica e Astronomia,\\
			Faculdade de Ciencias da Universidade do Porto, Porto 4169-007, Portugal}
\author{Carlo Heissenberg}
\affiliation{Institut de Physique Th\'eorique, CEA Saclay, CNRS, Universit\'e Paris-Saclay, 
			F-91191, Gif-sur-Yvette Cedex, France}
\author{Julio Parra-Martinez}
\affiliation{Institut des Hautes \'Etudes Scientifiques, F-91440 Bures-sur-Yvette, France}

\begin{abstract} 
We present the first computation of the nonlinear gravitational memory waveform for the scattering of two compact objects in General Relativity at leading order in the post-Minkowskian expansion. We use the scattering-amplitudes-based
representation of the gravitational waveform, which naturally expresses the nonlinear memory as the contribution of soft gravitons
emitted by the gravitational waves themselves. We perform the calculation by applying a multipolar decomposition to the waveform and using the reverse unitarity method to  obtain explicit exact-in-velocity predictions. We validate the
results by calculating the corresponding velocity-expanded post-Newtonian multipoles, finding perfect agreement. 
Our results complete the knowledge of the gauge-invariant non-analytic-in-frequency part of the $\mathcal{O}(G^3)$ multipolar waveform,
thus providing a useful benchmark for future calculations of this quantity.

\end{abstract}

\maketitle

\textit{Introduction.}--
The advent of gravitational-wave (GW) science has sparked the need for precision in gravitational physics. The success of the scientific program of current and future GW experiments hinges on our ability to extract accurate predictions for the dynamics and radiation of compact binary systems from General Relativity (GR). This is a nontrivial task, as Einstein's theory is highly nonlinear, and in order to solve it one must resort to numerical methods \cite{Pretorius:2005gq,Gourgoulhon:2012ffd} or various analytical approximations schemes, such as the post-Newtonian (PN) \cite{Blanchet:2013haa} and self-force \cite{Barack:2018yvs} theories. In recent years, a novel synergy between ideas from scattering amplitudes, effective field theory and classical GR has brought in a wealth of results and significantly advanced our ability to make predictions in the post-Minkowskian (PM) expansion \cite{Damour:2017zjx,Cheung:2018wkq,Bern:2019nnu,KoemansCollado:2019ggb,Bern:2019crd,Damour:2019lcq,Cristofoli:2020uzm,Damour:2020tta,Bern:2021dqo,Herrmann:2021lqe,DiVecchia:2021ndb,DiVecchia:2021bdo,Herrmann:2021tct,Bjerrum-Bohr:2021vuf,Bjerrum-Bohr:2021din,Dlapa:2021npj,Brandhuber:2021eyq,Bern:2021yeh,Dlapa:2021vgp,Manohar:2022dea,Bern:2022jvn,Dlapa:2022lmu,Barack:2023oqp,Dlapa:2023hsl,Damgaard:2023ttc,Cheung:2023lnj,Driesse:2024xad,Bern:2024adl,Cheung:2024byb,Driesse:2024feo,Heissenberg:2025ocy}, which applies for weak fields and relativistic velocities, employing Newton's constant, $G$, as an expansion parameter.

The main quantity measured in experiments is the waveform of the gravitational wave passing through the detector, $w_{\mu\nu}$. This is simply the radiative part of the gravitational field, $h_{\mu\nu}=g_{\mu\nu} - \eta_{\mu\nu}$, at distances $r\to\infty$ far from the emitter,
$ 
	h_{\mu\nu} \sim \frac{4G}{r}\,w_{\mu\nu}(T,n)\,, 
$ 
and a function of the observation time, $T$, and a null vector, $n^\mu$, characterizing the observer's direction. Following pioneering papers based on classical methods \cite{Peters:1964zz,Kovacs:1977uw,Kovacs:1978eu}, the gravitational waveform has been recently connected to scattering-amplitude and diagrammatic methods \cite{Kosower:2018adc,Jakobsen:2021smu,Mougiakakos:2021ckm,DiVecchia:2021bdo,Cristofoli:2021vyo,Caron-Huot:2023vxl,Biswas:2024ept} through $\mathcal{O}(G^2)$ \cite{Brandhuber:2023hhy,Herderschee:2023fxh,Elkhidir:2023dco,Georgoudis:2023lgf,Caron-Huot:2023vxl,Georgoudis:2023eke,Georgoudis:2023ozp,Bini:2023fiz,Georgoudis:2024pdz,Bini:2024rsy} leading to new results.

Perhaps the conceptually simplest observable feature of the waveform is the so-called \emph{memory} of the gravitational wave, which is the net change of the gravitational waveform during the observation process \cite{Zeldovich:1974gvh,Braginsky:1985vlg,Braginsky:1987kwo, Christodoulou:1991cr,Thorne:1992sdb}, 
\begin{equation}\label{eq:memT}
	\Big(
	\lim_{T\to\infty} w_{\mu\nu}(T,n)
	\Big) 
	- 
	\Big(
	\lim_{T\to-\infty}w_{\mu\nu}(T,n)
	\Big) 
    = - F_{\mu\nu}
    \,.
\end{equation}
The memory dictates a permanent displacement of free-falling test masses after the passing of the gravitational wave \cite{Braginsky:1985vlg, Braginsky:1987kwo} and is related by Fourier transform to a $1/\omega$ pole in the small-frequency limit $\omega\to0$ of the spectral waveform \cite{Strominger:2014pwa}. See \cite{Lasky:2016knh, Hubner:2019sly,Boersma:2020gxx,Grant:2022bla, Goncharov:2023woe,Gasparotto:2023fcg,Inchauspe:2024ibs} for prospects on its measurement at interferometers.

Scattering amplitudes also enjoy general factorization properties in the low-frequency limit known as soft theorems, and their leading-order incarnation in the context of gravity is the celebrated Weinberg soft theorem \cite{Weinberg:1964ew,Weinberg:1965nx}. As it turns out, the Weinberg soft factor \emph{is} the memory effect \cite{Strominger:2014pwa}, and similar general results also hold for subleading $\log\omega$ and $\omega(\log\omega)^2$ contributions in the low-frequency limit  \cite{Saha:2019tub,Sahoo:2021ctw} (see \cite{Alessio:2024onn,Fucito:2024wlg} for an all-order conjecture). Such soft theorems are nonperturbative in $G$ and thus provide all-order constraints for the PM expansion of the waveform. In Refs.~\cite{Jakobsen:2021smu,Mougiakakos:2021ckm} and \cite{Aoude:2023dui,Georgoudis:2023eke,Bini:2024rsy} it was checked that amplitude-based calculations are in perfect agreement with such constraints at $\mathcal{O}(G)$ and $\mathcal{O}(G^2)$.

The key idea at the basis of \cite{Weinberg:1964ew,Weinberg:1965nx}  is that the leading soft term is sourced by gravitons attached to the external lines of the process, to which the graviton couples in a democratic way owing to the equivalence principle. As already emphasized by Weinberg, the leading soft theorem is completely universal and includes the contributions due to soft gravitons emitted by \emph{other}, hard, gravitons. In the classical GR context, this contribution is known as \emph{nonlinear} memory \cite{Christodoulou:1991cr,Thorne:1992sdb}, as opposed to \emph{linear} memory sourced by the massive bodies \cite{Zeldovich:1974gvh,Braginsky:1987kwo}.

In this work we show how the nonlinear memory emerges from the amplitude representation of the gravitational waveform for the scattering of two compact objects, at the leading order at which it appears, $\mathcal{O}(G^3)$. Furthermore, we explicitly calculate the associated multipoles in the post-Minkowskian limit for the first time. 
This completes the predictions given by classical soft theorems for the non-analytic-in-$\omega$ part of the waveform at $\mathcal{O}(G^3)$, which includes the memory and logarithms of $\omega$.

\textit{Linear and nonlinear memory waveform.}--
The memory effect is captured by a simple pole in the low-frequency expansion of the spectral waveform $\tilde{w}_{\mu\nu}(\omega n)$ \cite{Zeldovich:1974gvh,Braginsky:1987kwo,Christodoulou:1991cr,Thorne:1992sdb,Strominger:2014pwa,Saha:2019tub,Sahoo:2021ctw}
\begin{equation}\label{eq:WeinbergPole}
	\tilde{w}_{\mu\nu}
	\sim
	-\frac{i}{\omega}\, F_{\mu\nu} + \mathcal{O}(\log\omega)\,,\text{ as }\omega\to0\,,
\end{equation}
 and affords the simple expression \cite{Weinberg:1964ew,Weinberg:1965nx,Saha:2019tub,Sahoo:2021ctw}
 \begin{equation}\label{eq:WeinbergF}
 	F^{\mu\nu} = \sum_{a} \frac{p_a^\mu p_a^\nu}{p_a\cdot n}\,.
 \end{equation}
In the Weinberg factor \eqref{eq:WeinbergF}, the sum encompasses \emph{all} hard external states that can emit soft gravitons. For a gravitational two-body scattering, this includes not only the two massive compact objects, with initial  momenta $-p_{1,2}^\mu = m_{1,2} v_{1,2}^\mu$ (and final momenta $p_{4,3}^\mu$), but also the dynamically generated gravitons themselves, which build the GW signal. Therefore, it is convenient to split $F^{\mu\nu} = f^{\mu\nu} + \delta F^{\mu\nu}$ in terms of 
\begin{equation}\label{eq:FdeltaF}
	f^{\mu\nu} = \sum_{a=1}^4 \frac{p_a^\mu p_a^\nu}{p_a\cdot n}\,,\qquad
	\delta F^{\mu\nu} = \int \widetilde{dk}\, \rho(k)\,\frac{k^\mu k^\nu}{k\cdot n}\,,
\end{equation}
where we introduced a shorthand notation for the integral over the Lorentz-invariant phase space of emitted gravitons, ($t^\mu$ being a timelike vector specifying the time direction)
\begin{equation}
\int \widetilde{dk} = \int \frac{d^Dk}{(2\pi)^{D-1}} \,  \theta(-t\cdot k)\,\delta(k^2)\,,
\end{equation}
and $\rho(k)$ denotes the corresponding phase-space distribution,
\begin{equation}\label{eq:rhoexpl}
	\rho = 8\pi G\,\tilde{w}_{\mu\nu} \left( \eta_{\mu\rho}\eta_{\nu\sigma} - \tfrac{1}{D-2}\,\eta_{\mu\nu}\eta_{\rho\sigma}\right) \tilde{w}^\ast_{\rho\sigma}\,.
\end{equation}
Since $f^{\mu\nu}$ is due to the field generated by the incoming and outgoing massive objects, it is referred to as \emph{linear} memory, while $\delta F^{\mu\nu}$ is due to soft gravitons emitted by hard gravitons, a phenomenon known as \emph{nonlinear} memory. Importantly,  gauge invariance  holds only for the sum of linear and nonlinear memory, $n_\nu F^{\mu\nu}=0$, thanks to 
\begin{equation}\label{eq:balancelaw}
\sum_{a=1}^4 p_a^\mu + \int \widetilde{dk}\,\rho(k)\,k^\mu = 0\,,
\end{equation}
which is the balance law linking mechanical and radiated energy-momentum.

While the expressions \eqref{eq:FdeltaF} are formally exact, they rely on the knowledge of the incoming and outgoing momenta of the massive objects, as well as on the full gravitational wave spectrum, $\rho(k)$. However, order by order in the PM expansion, we can evaluate them explicitly as functions of the initial velocities $v_{1,2}^{\mu}$, thus in particular of the Lorentz factor $\sigma=-v_1\cdot v_2 = 1/\sqrt{1-v^2} $, and of the impact parameter  $b^\mu$ orthogonal to them, $b\cdot v_{1,2}=0$.

\textit{Linear memory.}-- Before moving on to the more interesting case of nonlinear memory, let us recall the PM expansion of the linear memory contribution. This is determined by the impulses \cite{Cheung:2018wkq,Bern:2019nnu,KoemansCollado:2019ggb,Bern:2019crd,Cristofoli:2020uzm,Damour:2020tta,Herrmann:2021lqe,DiVecchia:2021ndb,Herrmann:2021tct}
\begin{subequations}
\begin{align}
    p_1^\mu + p_4^\mu 
    &= - Q\,\frac{b_e^\mu}{b_e} -  Q_{\parallel}\,\check{v}_2^\mu + \mathcal{O}(G^4)\,,\\
    p_2^\mu + p_3^\mu 
    &= + Q\,\frac{b_e^\mu}{b_e} -  Q_{\parallel}\,\check{v}_1^\mu + \mathcal{O}(G^4)\,,
\end{align}
\end{subequations}
where $\check{v}_{1,2}^\mu = (\sigma v_{2,1}^\mu - v_{1,2}^\mu) /(\sigma^2-1)$ and
\begin{equation}\label{eq:beb}
    b_e^\mu = b^\mu - \left(
	\frac{\check{v}_1^\mu}{2m_1}
	-
	\frac{\check{v}_2^\mu}{2m_2}
	\right)
	Q\,b_e\,.
\end{equation}
Explicit expressions for the transverse, $Q$, and longitudinal, $Q_{\parallel}$, components of the impulse as functions of $\sigma$ and $b$ can be found in e.g.~\cite{DiVecchia:2023frv}. Note that only the longitudinal component enters the balance law \eqref{eq:balancelaw} at $\mathcal{O}(G^3)$, which holds thanks to \cite{Herrmann:2021lqe}
\begin{equation}
    \boldsymbol{P}^\mu = \int \widetilde{dk}\,\rho(k)\,k^\mu = Q_{\parallel}\,(\check{v}_1^\mu + \check{v}_2^\mu)+ \mathcal{O}(G^4)\,.  
\end{equation}
Contracting with a fixed-helicity polarization vector $\varepsilon$ and expanding for small $G$ one thus finds
\begin{equation}\label{eq:S1S3T3}
	\varepsilon_\mu\,f^{\mu\nu} \varepsilon_\nu
	=
	Q\, S_1^{(\varepsilon)} + Q^3\, S_3^{(\varepsilon)} + Q_{\parallel}\,T_3^{(\varepsilon)} + \mathcal{O}(G^4)\,,
\end{equation}
in terms of elementary tensor structures $S^{(\varepsilon)}_{1,2}$ and $T^{(\varepsilon)}_3$ (see Eqs.~(2.62)--(2.64) of \cite{Alessio:2024onn} for explicit expressions).

\textit{Nonlinear memory from scattering amplitudes.}-- 
The gravitational waveform, $\tilde w^{\mu\nu}$, can be expressed in terms of scattering amplitudes, which are the $T$-matrix elements with $S =1 + iT$, as the impact-parameter Fourier transform of $W^{\mu\nu}/\sqrt{8\pi G}$
with \cite{Kosower:2018adc,Cristofoli:2021vyo,Brandhuber:2023hhy,Herderschee:2023fxh,Elkhidir:2023dco,Georgoudis:2023lgf,Georgoudis:2023ozp} 
\begin{equation}\label{eq:waveformKMOC}
\begin{split}
	W^{\mu\nu} &= 
	\begin{gathered}
	\begin{tikzpicture}[scale=.55]
		\draw[red,decorate,decoration={coil,,aspect=0,segment length=1.5mm,amplitude=.5mm,pre length=0pt,post length=0pt}] (0,0)  -- (2.05,0);
		\draw [ultra thick, blue] (-2,1.3) -- (0,0);
		\draw [ultra thick, blue] (0,0) -- (2,1.3);
		\draw [ultra thick, green!70!black] (-2,-1.3) -- (0,0);
		\draw [ultra thick, green!70!black] (0,0) -- (2,-1.3);
		\filldraw[black!10!white, thick] (0,0) ellipse (1 and 1);
		\filldraw[pattern=north west lines, thick] (0,0) ellipse (1 and 1);
		\draw[thick] (0,0) ellipse (1 and 1);
	\end{tikzpicture}
	\end{gathered}
	-
	i
	\sum_X
    \!\!
	\begin{gathered}
		\begin{tikzpicture}[scale=.65]
			\draw [ultra thick, blue] (-3,1.3) -- (-1.5,0);
			\draw [ultra thick, blue] (-1.5,0) .. controls (-1,1.5) and (1,1.5) .. (1.5,0);
			\draw [ultra thick, blue] (3,1.3) -- (1.5,0);\draw[red,decorate,decoration={coil,,aspect=0,segment length=1.5mm,amplitude=.5mm,pre length=0pt,post length=0pt}] (-1.5,0)  -- (-.22,0); 
			\draw[ultra thick] (-1.5,0)  .. controls (-1,-.9) and (1,-.9) .. (1.5,0);
			\draw [ultra thick, green!70!black] (-3,-1.3) -- (-1.5,0);
			\draw [ultra thick, green!70!black] (-1.5,0) .. controls (-1,-1.5) and (1,-1.5) .. (1.5,0);
			\draw [ultra thick, green!70!black] (3,-1.3) -- (1.5,0);
			\filldraw[black!10!white, thick] (-1.5,0) ellipse (.7 and .7);
			\filldraw[pattern = north west lines, thick] (-1.5,0) ellipse (.7 and .7);
			\draw[thick] (-1.5,0) ellipse (.7 and .7);
			\filldraw[black!10!white, thick] (1.5,0) ellipse (.7 and .7);
			\filldraw[pattern = north west lines, thick] (1.5,0) ellipse (.7 and .7);
			\draw[thick] (1.5,0) ellipse (.7 and .7);
			\node at (-.2,-.1)[below]{${}_X$};
			\draw [red] (0,-1.5) -- (0,1.5);
		\end{tikzpicture}
	\end{gathered}
    \end{split}
\end{equation}
and in the loop expansion $W^{\mu\nu} = \sum_{L} W^{\mu\nu}_L$.
To make the right-hand side of \eqref{eq:waveformKMOC} more explicit and expose its behavior in the classical limit, it is convenient to relate it to the $N$-matrix elements, with \cite{Damgaard:2021ipf} $S = e^{iN}$. At tree level, $W_0^{\mu\nu}$ coincides with the $2\to3$ amplitude \cite{Goldberger:2016iau,Luna:2017dtq}, while at one loop one finds 
\cite{Caron-Huot:2023vxl,Aoude:2023dui}
\begin{align}\label{eq:W1soft}
	&W_1^{\mu\nu}
	\sim
	\begin{gathered}
		\begin{tikzpicture}[scale=.55]
    \draw[red,decorate,decoration={coil,,aspect=0,segment length=1.5mm,amplitude=.5mm,pre length=0pt,post length=0pt}] (0,0)  -- (2.05,0);
			\draw [ultra thick, blue] (-2,1.3) -- (0,0);
			\draw [ultra thick, blue] (0,0) -- (2,1.3);
			\draw [ultra thick, green!70!black] (-2,-1.3) -- (0,0);
			\draw [ultra thick, green!70!black] (0,0) -- (2,-1.3);
			\filldraw[black!20!white, thick] (0,0) ellipse (1 and 1);
			\draw[thick] (0,0) ellipse (1 and 1);
			\filldraw[white, thick] (0,0) ellipse (.4 and .4);
			\draw[thick] (0,0) ellipse (.4 and .4);
			\node at (0,1)[above]{$N$};
			\node at (0,-1)[below]{$\phantom{N}$};
		\end{tikzpicture}
	\end{gathered}
    \\
	&+ \frac{i}{2}
	\left(
	\begin{gathered}
		\begin{tikzpicture}[scale=.55]
			\draw [ultra thick, blue] (-3,1.3) -- (-1.5,0);
			\draw [ultra thick, blue] (-1.5,0) .. controls (-1,1.5) and (1,1.5) .. (1.5,0);
			\draw [ultra thick, blue] (3,1.3) -- (1.5,0);
			\draw[red,decorate,decoration={coil,,aspect=0,segment length=1.5mm,amplitude=.5mm,pre length=0pt,post length=0pt}] (1.5,0)  -- (3,0);
			\draw [ultra thick, green!70!black] (-3,-1.3) -- (-1.5,0);
			\draw [ultra thick, green!70!black] (-1.5,0) .. controls (-1,-1.5) and (1,-1.5) .. (1.5,0);
			\draw [ultra thick, green!70!black] (3,-1.3) -- (1.5,0);
			\filldraw[black!20!white, thick] (-1.5,0) ellipse (.7 and .7);
			\draw[thick] (-1.5,0) ellipse (.7 and .7);
			\filldraw[black!20!white, thick] (1.5,0) ellipse (.7 and .7);
			\draw[thick] (1.5,0) ellipse (.7 and .7);
		\end{tikzpicture}
	\end{gathered}
	-
	\begin{gathered}
		\begin{tikzpicture}[scale=.55]
			\draw [ultra thick, blue] (-3,1.3) -- (-1.5,0);
			\draw [ultra thick, blue] (-1.5,0) .. controls (-1,1.5) and (1,1.5) .. (1.5,0);
			\draw [ultra thick, blue] (3,1.3) -- (1.5,0);\draw[red,decorate,decoration={coil,,aspect=0,segment length=1.5mm,amplitude=.5mm,pre length=0pt,post length=0pt}] (-1.5,0)  -- (0,0);
			\draw [ultra thick, green!70!black] (-3,-1.3) -- (-1.5,0);
			\draw [ultra thick, green!70!black] (-1.5,0) .. controls (-1,-1.5) and (1,-1.5) .. (1.5,0);
			\draw [ultra thick, green!70!black] (3,-1.3) -- (1.5,0);
			\filldraw[black!20!white, thick] (-1.5,0) ellipse (.7 and .7);
			\draw[thick] (-1.5,0) ellipse (.7 and .7);
			\filldraw[black!20!white, thick] (1.5,0) ellipse (.7 and .7);
			\draw[thick] (1.5,0) ellipse (.7 and .7);
		\end{tikzpicture}
	\end{gathered}
	\right)\,,
  \nonumber
\end{align}
where we have dropped terms that do not contribute at leading order as $\omega\to 0$.
Refs.~\cite{Aoude:2023dui,Georgoudis:2023eke,Bini:2024rsy} checked that the leading $\mathcal{O}(G)$, and to subleading, $\mathcal{O}(G^2)$, waveforms thus obtained are in perfect agreement with the corresponding expansion of the term $Q\, S_1^{(\varepsilon)}$ in \eqref{eq:S1S3T3}. Therefore, they agree with the general prediction given by the soft theorem and are completely exhausted by linear memory.

At two-loop order, we find that the terms contributing at leading order as $\omega\to 0$ take the form
\begin{widetext}
\begin{equation}\label{eq:W2soft}
	\begin{split}
		&W^{\mu\nu}_2 
		\sim 
		\begin{gathered}
			\begin{tikzpicture}[scale=.6]
				\draw[red,decorate,decoration={coil,,aspect=0,segment length=1.5mm,amplitude=.5mm,pre length=0pt,post length=0pt}] (0,0)  -- (2.255,0);
				\draw [ultra thick, blue] (-2,1.3) -- (0,0);
				\draw [ultra thick, blue] (0,0) -- (2,1.3);
				\draw [ultra thick, green!70!black] (-2,-1.3) -- (0,0);
				\draw [ultra thick, green!70!black] (0,0) -- (2,-1.3);
				\filldraw[black!20!white, thick] (0,0) ellipse (1 and 1);
				\draw[thick] (0,0) ellipse (1 and 1);
				\filldraw[white, thick] (-.4,0) ellipse (.2 and .4);
				\draw[thick] (-.4,0) ellipse (.2 and .4);
				\filldraw[white, thick] (.4,0) ellipse (.2 and .4);
				\draw[thick] (.4,0) ellipse (.2 and .4);
				\node at (0,1)[above]{$N$};
				\node at (0,-1)[below]{$\phantom{N}$};
			\end{tikzpicture}
		\end{gathered}
		+\frac{i}{2}
		\left(
		\begin{gathered}
			\begin{tikzpicture}[scale=.6]
				\draw [ultra thick, blue] (-3,1.3) -- (-1.5,0);
				\draw [ultra thick, blue] (-1.5,0) .. controls (-1,1.5) and (1,1.5) .. (1.5,0);
				\draw [ultra thick, blue] (3,1.3) -- (1.5,0);
				\draw[red,decorate,decoration={coil,,aspect=0,segment length=1.5mm,amplitude=.5mm,pre length=0pt,post length=0pt}] (-1.5,0)  -- (2.88,0);
				\draw [ultra thick, green!70!black] (-3,-1.3) -- (-1.5,0);
				\draw [ultra thick, green!70!black] (-1.5,0) .. controls (-1,-1.5) and (1,-1.5) .. (1.5,0);
				\draw [ultra thick, green!70!black] (3,-1.3) -- (1.5,0);
				\filldraw[black!20!white, thick] (-1.5,0) ellipse (.7 and .7);
				\draw[thick] (-1.5,0) ellipse (.7 and .7);
				\filldraw[black!20!white, thick] (1.5,0) ellipse (.7 and .7);
				\draw[thick] (1.5,0) ellipse (.7 and .7);
			\end{tikzpicture}
		\end{gathered}
		-
		\begin{gathered}
			\begin{tikzpicture}[scale=.6]
				\draw [ultra thick, blue] (-3,1.3) -- (-1.5,0);
				\draw [ultra thick, blue] (-1.5,0) .. controls (-1,1.5) and (1,1.5) .. (1.5,0);
				\draw [ultra thick, blue] (3,1.3) -- (1.5,0);
				\draw[red,decorate,decoration={coil,,aspect=0,segment length=1.5mm,amplitude=.5mm,pre length=0pt,post length=0pt}] (-1.5,0)  -- (.13,0);
				\draw[red,decorate,decoration={coil,,aspect=0,segment length=1.5mm,amplitude=.5mm,pre length=0pt,post length=0pt}] (-1.5,0)  .. controls (-1,-.8) and (1,-.8) .. (1.5,0);
				\draw [ultra thick, green!70!black] (-3,-1.3) -- (-1.5,0);
				\draw [ultra thick, green!70!black] (-1.5,0) .. controls (-1,-1.5) and (1,-1.5) .. (1.5,0);
				\draw [ultra thick, green!70!black] (3,-1.3) -- (1.5,0);
				\filldraw[black!20!white, thick] (-1.5,0) ellipse (.7 and .7);
				\draw[thick] (-1.5,0) ellipse (.7 and .7);
				\filldraw[black!20!white, thick] (1.5,0) ellipse (.7 and .7);
				\draw[thick] (1.5,0) ellipse (.7 and .7);
			\end{tikzpicture}
		\end{gathered}
		\right)
		\\
		&
		+ \frac{i}{2}
		\left(
		\begin{gathered}
			\begin{tikzpicture}[scale=.4]
				\draw [ultra thick, blue] (-3,1.3) -- (-1.5,0);
				\draw [ultra thick, blue] (-1.5,0) .. controls (-1,1.5) and (1,1.5) .. (1.5,0);
				\draw [ultra thick, blue] (3,1.3) -- (1.5,0);
				\draw[red,decorate,decoration={coil,,aspect=0,segment length=1.5mm,amplitude=.5mm,pre length=0pt,post length=0pt}] (1.5,0)  -- (3.19,0);
				\draw [ultra thick, green!70!black] (-3,-1.3) -- (-1.5,0);
				\draw [ultra thick, green!70!black] (-1.5,0) .. controls (-1,-1.5) and (1,-1.5) .. (1.5,0);
				\draw [ultra thick, green!70!black] (3,-1.3) -- (1.5,0);
				\filldraw[black!20!white, thick] (-1.5,0) ellipse (.7 and .7);
				\draw[thick] (-1.5,0) ellipse (.7 and .7);
				\filldraw[black!20!white, thick] (1.5,0) ellipse (.7 and .7);
				\draw[thick] (1.5,0) ellipse (.7 and .7);
				\filldraw[white, thick] (1.5,0) ellipse (.3 and .3);
				\draw[thick] (1.5,0) ellipse (.3 and .3);
				\node at (1.5,.7)[above]{$N$};
			\end{tikzpicture}
		\end{gathered}
		+
			\begin{gathered}
			\begin{tikzpicture}[scale=.4]
				\draw [ultra thick, blue] (-3,1.3) -- (-1.5,0);
				\draw [ultra thick, blue] (-1.5,0) .. controls (-1,1.5) and (1,1.5) .. (1.5,0);
				\draw [ultra thick, blue] (3,1.3) -- (1.5,0);
				\draw[red,decorate,decoration={coil,,aspect=0,segment length=1.5mm,amplitude=.5mm,pre length=0pt,post length=0pt}] (1.5,0)  -- (3.19,0);
				\draw [ultra thick, green!70!black] (-3,-1.3) -- (-1.5,0);
				\draw [ultra thick, green!70!black] (-1.5,0) .. controls (-1,-1.5) and (1,-1.5) .. (1.5,0);
				\draw [ultra thick, green!70!black] (3,-1.3) -- (1.5,0);
				\filldraw[black!20!white, thick] (-1.5,0) ellipse (.7 and .7);
				\draw[thick] (-1.5,0) ellipse (.7 and .7);
				\filldraw[white, thick] (-1.5,0) ellipse (.3 and .3);
				\draw[thick] (-1.5,0) ellipse (.3 and .3);
				\filldraw[black!20!white, thick] (1.5,0) ellipse (.7 and .7);
				\draw[thick] (1.5,0) ellipse (.7 and .7);
				\node at (-1.5,.7)[above]{$N$};
			\end{tikzpicture}
		\end{gathered}
		-
		\begin{gathered}
			\begin{tikzpicture}[scale=.4]
				\draw [ultra thick, blue] (-3,1.3) -- (-1.5,0);
				\draw [ultra thick, blue] (-1.5,0) .. controls (-1,1.5) and (1,1.5) .. (1.5,0);
				\draw [ultra thick, blue] (3,1.3) -- (1.5,0);
				\draw[red,decorate,decoration={coil,,aspect=0,segment length=1.5mm,amplitude=.5mm,pre length=0pt,post length=0pt}] (-1.5,0)  -- (0.19,0);
				\draw [ultra thick, green!70!black] (-3,-1.3) -- (-1.5,0);
				\draw [ultra thick, green!70!black] (-1.5,0) .. controls (-1,-1.5) and (1,-1.5) .. (1.5,0);
				\draw [ultra thick, green!70!black] (3,-1.3) -- (1.5,0);
				\filldraw[black!20!white, thick] (-1.5,0) ellipse (.7 and .7);
				\draw[thick] (-1.5,0) ellipse (.7 and .7);
				\filldraw[black!20!white, thick] (1.5,0) ellipse (.7 and .7);
				\draw[thick] (1.5,0) ellipse (.7 and .7);
				\filldraw[white, thick] (1.5,0) ellipse (.3 and .3);
				\draw[thick] (1.5,0) ellipse (.3 and .3);
				\node at (1.5,.7)[above]{$N$};
			\end{tikzpicture}
		\end{gathered}
		-
		\begin{gathered}
			\begin{tikzpicture}[scale=.4]
				\draw [ultra thick, blue] (-3,1.3) -- (-1.5,0);
				\draw [ultra thick, blue] (-1.5,0) .. controls (-1,1.5) and (1,1.5) .. (1.5,0);
				\draw [ultra thick, blue] (3,1.3) -- (1.5,0);
				\draw[red,decorate,decoration={coil,,aspect=0,segment length=1.5mm,amplitude=.5mm,pre length=0pt,post length=0pt}] (-1.5,0)  -- (0.19,0);
				\draw [ultra thick, green!70!black] (-3,-1.3) -- (-1.5,0);
				\draw [ultra thick, green!70!black] (-1.5,0) .. controls (-1,-1.5) and (1,-1.5) .. (1.5,0);
				\draw [ultra thick, green!70!black] (3,-1.3) -- (1.5,0);
				\filldraw[black!20!white, thick] (-1.5,0) ellipse (.7 and .7);
				\draw[thick] (-1.5,0) ellipse (.7 and .7);
				\filldraw[white, thick] (-1.5,0) ellipse (.3 and .3);
				\draw[thick] (-1.5,0) ellipse (.3 and .3);
				\filldraw[black!20!white, thick] (1.5,0) ellipse (.7 and .7);
				\draw[thick] (1.5,0) ellipse (.7 and .7);
				\node at (-1.5,.7)[above]{$N$};
			\end{tikzpicture}
		\end{gathered}
		\right)
		\\
		&-\frac{1}{6}\left(
		\begin{gathered}
			\begin{tikzpicture}[scale=.4]
				\draw [ultra thick, blue] (-3,1.3) -- (-1.5,0);
				\draw [ultra thick, blue] (-1.5,0) .. controls (-1,1.5) and (1,1.5) .. (1.5,0);
				\draw [ultra thick, blue] (1.5,0) .. controls (2,1.5) and (4,1.5) .. (4.5,0);
				\draw [ultra thick, blue] (6,1.3) -- (4.5,0);
				\draw[red,decorate,decoration={coil,,aspect=0,segment length=1.5mm,amplitude=.5mm,pre length=0pt,post length=0pt}] (-1.5,0)  -- (.2,0);
				\draw [ultra thick, green!70!black] (-3,-1.3) -- (-1.5,0);
				\draw [ultra thick, green!70!black] (-1.5,0) .. controls (-1,-1.5) and (1,-1.5) .. (1.5,0);
				\draw [ultra thick, green!70!black] (1.5,0) .. controls (2,-1.5) and (4,-1.5) .. (4.5,0);
				\draw [ultra thick, green!70!black] (6,-1.3) -- (4.5,0);
				\filldraw[black!20!white, thick] (-1.5,0) ellipse (.7 and .7);
				\draw[thick] (-1.5,0) ellipse (.7 and .7);
				\filldraw[black!20!white, thick] (1.5,0) ellipse (.7 and .7);
				\draw[thick] (1.5,0) ellipse (.7 and .7);
				\filldraw[black!20!white, thick] (4.5,0) ellipse (.7 and .7);
				\draw[thick] (4.5,0) ellipse (.7 and .7);
			\end{tikzpicture}
		\end{gathered}
		-2
		\begin{gathered}
			\begin{tikzpicture}[scale=.4]
				\draw [ultra thick, blue] (-3,1.3) -- (-1.5,0);
				\draw [ultra thick, blue] (-1.5,0) .. controls (-1,1.5) and (1,1.5) .. (1.5,0);
				\draw [ultra thick, blue] (1.5,0) .. controls (2,1.5) and (4,1.5) .. (4.5,0);
				\draw [ultra thick, blue] (6,1.3) -- (4.5,0);
				\draw[red,decorate,decoration={coil,,aspect=0,segment length=1.5mm,amplitude=.5mm,pre length=0pt,post length=0pt}] (1.5,0)  -- (3.2,0);
				\draw [ultra thick, green!70!black] (-3,-1.3) -- (-1.5,0);
				\draw [ultra thick, green!70!black] (-1.5,0) .. controls (-1,-1.5) and (1,-1.5) .. (1.5,0);
				\draw [ultra thick, green!70!black] (1.5,0) .. controls (2,-1.5) and (4,-1.5) .. (4.5,0);
				\draw [ultra thick, green!70!black] (6,-1.3) -- (4.5,0);
				\filldraw[black!20!white, thick] (-1.5,0) ellipse (.7 and .7);
				\draw[thick] (-1.5,0) ellipse (.7 and .7);
				\filldraw[black!20!white, thick] (1.5,0) ellipse (.7 and .7);
				\draw[thick] (1.5,0) ellipse (.7 and .7);
				\filldraw[black!20!white, thick] (4.5,0) ellipse (.7 and .7);
				\draw[thick] (4.5,0) ellipse (.7 and .7);
			\end{tikzpicture}
		\end{gathered}
		+
		\begin{gathered}
			\begin{tikzpicture}[scale=.4]
				\draw [ultra thick, blue] (-3,1.3) -- (-1.5,0);
				\draw [ultra thick, blue] (-1.5,0) .. controls (-1,1.5) and (1,1.5) .. (1.5,0);
				\draw [ultra thick, blue] (1.5,0) .. controls (2,1.5) and (4,1.5) .. (4.5,0);
				\draw [ultra thick, blue] (6,1.3) -- (4.5,0);
				\draw[red,decorate,decoration={coil,,aspect=0,segment length=1.5mm,amplitude=.5mm,pre length=0pt,post length=0pt}] (4.5,0)  -- (6.2,0);
				\draw [ultra thick, green!70!black] (-3,-1.3) -- (-1.5,0);
				\draw [ultra thick, green!70!black] (-1.5,0) .. controls (-1,-1.5) and (1,-1.5) .. (1.5,0);
				\draw [ultra thick, green!70!black] (1.5,0) .. controls (2,-1.5) and (4,-1.5) .. (4.5,0);
				\draw [ultra thick, green!70!black] (6,-1.3) -- (4.5,0);
				\filldraw[black!20!white, thick] (-1.5,0) ellipse (.7 and .7);
				\draw[thick] (-1.5,0) ellipse (.7 and .7);
				\filldraw[black!20!white, thick] (1.5,0) ellipse (.7 and .7);
				\draw[thick] (1.5,0) ellipse (.7 and .7);
				\filldraw[black!20!white, thick] (4.5,0) ellipse (.7 and .7);
				\draw[thick] (4.5,0) ellipse (.7 and .7);
			\end{tikzpicture}
		\end{gathered}
		\right)
	\end{split}
\end{equation}
\end{widetext}

Using the Weinberg soft theorem for all the elementary building blocks, we have checked that \eqref{eq:W2soft} agrees with the sub-subleading, $\mathcal{O}(G^3)$, expansion of the general expression in \eqref{eq:WeinbergPole}, \eqref{eq:FdeltaF} given by the soft theorem.

More in detail, the $N$-matrix element in the first line of \eqref{eq:W2soft}  plus the second and third lines match the expansion of the linear memory terms $Q S_1^{\mu\nu} + Q^3 S_3^{\mu\nu}$ in \eqref{eq:S1S3T3}. Notice that, like at one loop, also $\tilde{w}_2$ given by \eqref{eq:W2soft} is expressed in terms of the vectors $v_1^\mu$, $v_2^\mu$ and $b^\mu$ characterizing the \textit{initial} state. 
Another point to keep in mind to simplify the comparison is that the $N$-matrix element for the $2\to2$ process is directly related to the 3PM radial action, \cite{Bern:2021dqo,Bjerrum-Bohr:2021din}, whose $J$ derivative gives the 3PM deflection angle.

The novelty is represented here by the appearance of the nonlinear memory contribution $\delta F^{\mu\nu}$. This emerges in the gauge invariant combination $\delta F^{\mu\nu} + Q_{\parallel}\,T_3^{\mu\nu}$ from the cuts within the round parentheses in the first line of \eqref{eq:W2soft}. To show this, it is convenient to use the reverse-unitarity representation of the nonlinear memory,
\begin{equation}\label{eq:Kernel}
	\delta F^{\mu\nu}
	=
	\operatorname{FT}_4
	\int_{q_1}
	\int_{k}
	\,\frac{k^\mu k^\nu}{k\cdot n}
	\begin{gathered}
		\begin{tikzpicture}[scale=.7]
			\draw [ultra thick, blue] (-3,1.3) -- (-1.5,0);
			\draw [ultra thick, blue] (-1.5,0) .. controls (-1,1.5) and (1,1.5) .. (1.5,0);
			\draw [ultra thick, blue] (3,1.3) -- (1.5,0);
			\draw[red,decorate,decoration={coil,,aspect=0,segment length=1.5mm,amplitude=.5mm,pre length=0pt,post length=0pt}] (-1.5,0)  -- (1.5,0);
			\draw [ultra thick, green!70!black] (-3,-1.3) -- (-1.5,0);
			\draw [ultra thick, green!70!black] (-1.5,0) .. controls (-1,-1.5) and (1,-1.5) .. (1.5,0);
			\draw [ultra thick, green!70!black] (3,-1.3) -- (1.5,0);
			\filldraw[black!20!white, thick] (-1.5,0) ellipse (.8 and .8);
			\draw[thick] (-1.5,0) ellipse (.8 and .8);
			\filldraw[black!20!white, thick] (1.5,0) ellipse (.8 and .8);
			\draw[thick] (1.5,0) ellipse (.8 and .8);
			\draw[->] (-.2,-.2)--(.2,-.2);
			\node at (0,0)[above]{$k$};
			\draw[->] (-.2,.95)--(.2,.95);
			\node at (0,1)[above]{$q_1-p_1$};
		\end{tikzpicture}
	\end{gathered}\,,
\end{equation}
as well as the analogous ones for $\boldsymbol{Q}_{1}^\alpha = -Q_\parallel \check{v}_2^\alpha$ (resp.~$\boldsymbol{Q}_{2}^\alpha = -Q_\parallel \check{v}_1^\alpha$), in which $k^\mu k^\nu/k\cdot n$ is replaced by a  factor $q_1^\alpha-\frac{1}{2}\,q^\alpha$ (resp.~$-q_1^\alpha +\frac{1}{2}\,q^\alpha - k^\alpha$), see e.g.~\cite{DiVecchia:2023frv}.
In \eqref{eq:Kernel}, we use the shorthand $\int_q = \int \frac{d^Dq}{(2\pi)^D}$, the Fourier transform
\begin{equation}
	\operatorname{FT}_4[f]
	=
    4\pi^2 
	\int_q  
	e^{ib\cdot q}\,
	\delta(2p_1\cdot q)\,
	\delta(2p_2\cdot q)
	f(q)\,,
\end{equation}
and the contraction between the two tensor structures as in \eqref{eq:rhoexpl} is implicit.
Additionally, all exposed lines are on-shell, so that the integrand includes the Lorentz-invariant phase-space measure $d\text{LIPS}(q_1,k) $, which in the near-forward limit reads
\begin{equation}
	d\text{LIPS}(q_1,k) = (2\pi)^3\delta(2p_1\cdot q_1) \delta(2p_2\cdot(q_1+k))\theta(k^0)\delta(k^2)\,.
\end{equation}
The integral appearing on the right-hand side of \eqref{eq:Kernel} can be interpreted as the three-particle cut of a two-loop integral and thus calculated from the corresponding discontinuity.
Evaluating \eqref{eq:Kernel} requires discussing families of two-loop master integrals that, in addition to the elastic case analyzed in \cite{Parra-Martinez:2020dzs,DiVecchia:2021bdo,Herrmann:2021tct,Bjerrum-Bohr:2021vuf,Bjerrum-Bohr:2021din,Brandhuber:2021eyq}, involve the extra factor $(k\cdot n)^{-1}$.

\textit{Multipoles from reverse-unitarity.}--
Generically, the nonlinear memory is a function of three dimensionless variables: two angles denoting the orientation in the celestial sphere and the relative speed of the incoming massive objects. Instead of working at fixed $n^\mu$, we first perform the multipole decomposition of $\delta F^{\mu\nu}$ in a reference frame characterized by the time direction $t^\mu$.
Introducing the spin-weighted spherical harmonics $Y_{\pm 2}^{\ell m}$ (see Appendix~\ref{app:SWSH} for more details), we thus consider
\begin{equation}\label{eq:decomposelm}
	\delta F^{\ell m} = \oint Y_{\pm 2}^{\ell m \ast}  \, \delta F_{\mu\nu} \,\varepsilon_{\pm}^\mu \varepsilon_{\pm}^\nu\, d\Omega\,.
\end{equation}
We find that the result for the both helicities is the same, i.e.~the $V$-multipoles vanish, and reads (see Refs.~\cite{Blanchet:1992br,Favata:2009ii,Favata:2008yd,Blanchet:2013haa} for an equivalent formula in terms of Cartesian multipoles)
\begin{equation}\label{eq:deltaFellm}
	\delta F^{\ell m}  
	= - \frac{2\pi}{\ell(\ell-1)} \, \mathcal{N}_2^{\ell m} Y^{\ell m \ast}_{\mu_1\cdots \mu_\ell} \operatorname{FT}_4 [\mathbb{I}^{\mu_1\cdots \mu_\ell}]\,,
\end{equation}
where $Y_{\mu_1\cdots \mu_\ell}^{\ell m}$ denotes the standard basis of symmetric trace-free (STF) tensors,
$\mathcal{N}_2^{\ell m}$ is a normalization factor (see Eq.~\eqref{eq:calNs} in Appendix~\ref{app:SWSH}) and
\begin{equation}\label{eq:BBI}
	\begin{split}
	\mathbb{I}^{\mu_1\cdots \mu_\ell}
	&=
	\int_{q_1}\int_k
	\frac{k^{\mu_1}\cdots k^{\mu_\ell}}{(-t\cdot k)^{\ell-1}}
	\begin{gathered}
		\begin{tikzpicture}[scale=.7]
			\draw [ultra thick, blue] (-3,1.3) -- (-1.5,0);
			\draw [ultra thick, blue] (-1.5,0) .. controls (-1,1.5) and (1,1.5) .. (1.5,0);
			\draw [ultra thick, blue] (3,1.3) -- (1.5,0);
			\draw[red,decorate,decoration={coil,,aspect=0,segment length=1.5mm,amplitude=.5mm,pre length=0pt,post length=0pt}] (-1.5,0)  -- (1.5,0);
			\draw [ultra thick, green!70!black] (-3,-1.3) -- (-1.5,0);
			\draw [ultra thick, green!70!black] (-1.5,0) .. controls (-1,-1.5) and (1,-1.5) .. (1.5,0);
			\draw [ultra thick, green!70!black] (3,-1.3) -- (1.5,0);
			\filldraw[black!20!white, thick] (-1.5,0) ellipse (.8 and .8);
			\draw[thick] (-1.5,0) ellipse (.8 and .8);
			\filldraw[black!20!white, thick] (1.5,0) ellipse (.8 and .8);
			\draw[thick] (1.5,0) ellipse (.8 and .8);
			\draw[->] (-.2,-.2)--(.2,-.2);
			\node at (0,0)[above]{$k$};
			\draw[->] (-.2,.95)--(.2,.95);
			\node at (0,1.1)[above]{$q_1-p_1$};
		\end{tikzpicture}
	\end{gathered}
	\end{split}
\end{equation}
where $-t\cdot k$ in the denominator is the frequency of the graviton with four-momentum $k^\mu$ as measured in the chosen reference frame.

To calculate the nonlinear memory multipoles, we choose to align the axes of our reference frame as follows, 
\begin{equation}\label{eq:txyrefs}
	t^\mu = v_1^\mu\,,\qquad
 	\hat{x}^\mu=\frac{b^{\mu}}{b}  \,,\qquad
	\hat{y}^\mu = \check{v}_{2}^\mu\,\sqrt{\sigma^2-1} \,,
\end{equation}
so that the $Y^{\ell m}_{\mu_1\cdots \mu_\ell}$ are purely spatial and can be expressed in terms of the characteristic vectors of the problems. For instance, we have 
\begin{subequations}
\begin{align}
Y^{2(\pm2)}_{i_1 i_2}
&=\frac{1}{4}\sqrt{\frac{15}{2\pi}}
\, (\hat x\pm i \hat y)_{i_1} (\hat x\pm i \hat y)_{i_2}\,,
\\
Y^{20}_{i_1 i_2}
&=\frac{3}{4}\sqrt{\frac{5}{\pi}}
\, (\hat{z}_{i_1} \hat{z}_{i_2}-\tfrac{1}{3}\,\delta_{i_1i_2})\,,
\end{align}
\end{subequations}
for the structures relevant to the explicit results displayed in Table~\ref{table:l2} below. Thus, the resulting integrals in Eq.~\eqref{eq:BBI} fall in the families described in \cite{Parra-Martinez:2020dzs,DiVecchia:2021bdo,Herrmann:2021tct,Bjerrum-Bohr:2021vuf,Bjerrum-Bohr:2021din,Brandhuber:2021eyq}, with the addition of extra powers of the propagator $-u_1\cdot k$. Using  integration-by-parts (IBP) identities, these can be easily reduced  to a set of master integrals, which are the same as those in \cite{Parra-Martinez:2020dzs} plus an additional integral in the ladder sector.  Computing the IBP reduction of higher multipoles, while systematic, becomes computationally intensive because the tensor rank of the numerator increases linearly with the multipole number $\ell$. Finally, we calculate the master integrals using the by-now-familiar method of canonical differential equations in the soft region \cite{Parra-Martinez:2020dzs,DiVecchia:2021bdo,Herrmann:2021tct} (more details will be provided in \cite{toapAGVGCHJPP}).

\textit{Results.}--
The resulting multipoles of the leading-order nonlinear memory take the form
\begin{equation}\label{eq:deltaFlmexplicit}
	\delta F^{\ell m}
	= \frac{G^3\pi^2m_1^2 m_2^2}{b^3 (\sigma^2-1)^{\frac{3}{2}}}\,i^{\ell}\mathcal{N}_2^{\ell m}{\cal F}^{\ell m}(\sigma) + {\cal O}(G^4)\,,
\end{equation}
where we introduced the function of velocity
\begin{equation}\label{eq:calFdefinition}
	\mathcal{F}^{\ell m}(\sigma)
	= 
	\frac{f^{\ell m}_1
	+
	f_2^{\ell m}\log\left(\tfrac{\sigma+1}{2}\right)
	+
	f_3^{\ell m}\,
	\tfrac{\operatorname{arccosh}\sigma}{\sqrt{\sigma^2-1}}}{(\sigma^2-1)^{\frac{\ell}{2}}}\,,
\end{equation}
which depends on the multipolar numbers. The coefficients $f_i^{\ell m}$ are real polynomials in $\sigma$, which are provided in Table~\ref{table:l2} for the quadrupolar ($\ell = 2$) waveform, and 
in Appendix~\ref{app:polynomials} up to $\ell=5$. 
We note that $\delta F^{\ell m} = 0$ if $\ell+m$ is odd and, as a consequence of the symmetry of $\rho(k)$ under $b\mapsto -b$ at leading order, $\delta F^{\ell(-m)} = (-)^\ell \delta F^{\ell m}$. We emphasize that these results and the ones in Appendix~\ref{app:polynomials} hold in the rest frame of particle 1 (see Eq.~\eqref{eq:txyrefs}), and thus only coincide with those calculated in the center-of-mass frame to leading order in the mass-ratio $m_2/m_1\ll1$.

\begin{table}[H]
\setlength{\tabcolsep}{1pt} 
\renewcommand{\arraystretch}{3}
\begin{tabular}{|c|}
\hline
\scalebox{0.84}{\tabeq{10cm}{
 f_1^{22} &= \frac{1}{960} (-4200 \sigma ^8+92820 \sigma ^7
		-124980 \sigma ^6-133050 \sigma ^5+106492 \sigma ^4
		\\
		&\quad+489510 \sigma ^3-627644 \sigma ^2+200025 \sigma +5212)
		\\ 
		f_2^{22} &=\frac{1}{8} (\sigma^2-1)^2\left(35 \sigma ^4+420 \sigma ^3-510 \sigma ^2+292 \sigma +67\right) \\
		f_3^{22} &= \frac{1}{64} (-280 \sigma ^9-2880 \sigma ^8+2380 \sigma ^7+9072 \sigma ^6-4996 \sigma ^5-11790 \sigma ^4
		\\
		&\quad+3460 \sigma ^3+7008 \sigma ^2-564 \sigma -1689) 
        \\
        f_1^{20} &= \frac{1}{2880}(-4200 \sigma ^8-176568 \sigma ^7+199020 \sigma ^6+545268 \sigma ^5-782780 \sigma ^4
			\\
			&\quad-424458 \sigma ^3+757396 \sigma ^2+52923 \sigma -175196)
		\\ 
		f_2^{20} &=\frac{1}{24} (\sigma^2-1)(35 \sigma ^6-1230 \sigma ^5+385 \sigma ^4+2188 \sigma ^3  \\
        &\quad-3419 \sigma ^2+1538 \sigma +119) \\
		f_3^{20} &=\frac{1}{192} (-280 \sigma ^9+10320 \sigma ^8-5060 \sigma ^7-29256 \sigma ^6+14516 \sigma ^5
        \\&\quad+37476 \sigma ^4 -10748 \sigma ^3-24348 \sigma ^2+1956 \sigma +5997 )
}}\\
\hline
\end{tabular}
\caption{Functions specifying the quadrupolar nonlinear memory waveform.}
\label{table:l2}
\end{table}

For completeness, let us also provide here the explicit multipole decomposition for the ``longitudinal'' component $T_{3}^{\mu\nu}$ of the linear memory, which appears in the gauge-invariant combination $\delta F^{\mu\nu} + Q_{\parallel} T_3^{\mu\nu}$ together with the nonlinear memory:
\begin{equation}\label{eq:multipolesT3}
	Q_{\parallel}\,T_{3}^{\ell m} = 
	\frac{\partial a^{\ell m}(\sigma)}{\partial \sigma}\,Q_{\parallel}\,,
\end{equation}
where 
\begin{equation}\label{eq:fLinMen}
a^{\ell m}(\sigma)
=
	i^m\pi \sigma\, v^{\ell}\,\ell!\,a^0_{\ell m} \,{}_2 F_{1}\left(\tfrac{\ell-1}{2},\tfrac{\ell}{2},\ell+\tfrac{3}{2},v^{2}\right).
\end{equation}
See Eq.~\eqref{eq:alllinmemmultipoles} in Appendix~\ref{app:multipolelinear} for an analogous formula that computes all linear memory multipoles.

General results exist beyond the $1/\omega$ pole and also fix the $\log\omega$ and $\omega(\log\omega)^2$ terms in the small-frequency expansion \cite{Sahoo:2018lxl,Saha:2019tub}. It was observed that nonlinear contributions, due to gravitons, to such terms admit a simple rewriting \cite{Sahoo:2021ctw} in terms of the total initial and final four-momentum of the massive objects. In particular, this makes it manifest that nonlinear effects in  
$\log\omega$ and $\omega(\log\omega)^2$ only appear from 
$\mathcal{O}(G^4)$. Moreover, recently a conjecture for the resummation of all leading logs of the form $\omega^{n-1}(\log\omega)^n$ for $n\ge3$ was proposed based on checks in the small-velocity limit \cite{Alessio:2024onn} and small-mass-ratio limit \cite{Fucito:2024wlg}. 
It is straightforward to compute the associated multipoles through $\mathcal{O}(G^3)$ in a generic frame (see Appendix~\ref{app:multipolelinear} for more details). 
Finally, very recent results also encompass multipoles due to ``UV logs'' such as $\omega\log\omega$ \cite{Blanchet:1997jj,Goldberger:2009qd,Fucito:2024wlg,Ivanov:2025ozg}.

Our results for the nonlinear memory multipoles thus complete this picture by providing the missing non-analytic terms of the waveform at ${\cal O}(G^3)$.

\textit{Checks.}--
As an independent way of calculating the nonlinear memory waveform, in the small-velocity expansion, one can employ the explicit expression for the tree-level and one-loop waveform calculated in Refs.~\cite{Georgoudis:2024pdz,Bini:2024rsy} and substitute them in \eqref{eq:rhoexpl}. The resulting angular integrals are then elementary when performed order by order in the velocity. In Ref.~\cite{toapCHRR}, this alternative strategy has been employed to obtain the nonlinear memory waveform $\delta F^{\mu\nu}$ at $\mathcal{O}(G^3)$ up to $\mathcal{O}(v^{19})$ for small $v$, i.e.~9PN relative to the leading order, which scales like $\mathcal{O}(v)$. After performing the multipole decomposition \eqref{eq:decomposelm}, we find complete agreement between those results and the small-velocity expansion of the present results.  Moreover, we find agreement with the previous results of Refs.~\cite{Wiseman:1991ss} and \cite{Favata:2011qi}, which provide the leading result for small $v$ to $\mathcal{O}(G^3)$ and to all orders in $G m/(v^2 b)$, respectively.

It is worth mentioning that, in the scattering scenario considered here, the nonlinear memory contribution $\delta F^{\mu\nu}$ appears at $\mathcal{O}(G^3 v)$, i.e.~2.5 orders further suppressed in the PN expansion compared to the leading quadrupole sitting in $f^{\mu\nu}$ \cite{Wiseman:1991ss}.

\textit{Conclusions.}--
In this work we have analyzed the nonlinear memory waveform for the scattering of compact objects in GR showing how it emerges from scattering amplitudes. For the first time, using the method of reverse unitarity \cite{Anastasiou:2002qz,Anastasiou:2003yy,Anastasiou:2015yha,Herrmann:2021lqe}, we have computed the multipoles of the nonlinear memory at leading order in the post-Minkowskian expansion, i.e., at ${\cal O}(G^3)$ and all-orders in velocity. By taking this step, we have completed our knowledge of the part of the ${\cal O}(G^3)$ multipolar waveform which is non-analytic in frequency, which should serve as a useful check for future calculations of this quantity.

There is, nonetheless, more to say about the nonlinear memory. The way in which it arose in our calculation from the soft limit of a $2\to 4$ amplitude reveals that, contrary to previous expectations \cite{Britto:2021pud,Cristofoli:2021jas}, classical gravitational radiation is not completely described by a (non-squeezed) coherent state. This, for instance, implies that incorporating the nonlinear memory in the eikonal framework will require going beyond the proposals in \cite{Cristofoli:2021jas,DiVecchia:2022piu} (see also \cite{Aoude:2024jxd,Alessio:2025flu}).  Furthermore, this suggests there might exist  classical correlations between the gravitational-wave signals at various detectors, arising from the nonlinear nature of GR, which could serve as a novel test of Einstein's theory. 

\textit{Acknowledgments.}--
We thank Rodolfo Russo for collaboration in the early stages of this work and on related projects, and Gabriele Veneziano for insightful observations on the relation between linear and nonlinear memory. We are also grateful to Donato Bini,  Luc Blanchet, Thibault Damour, Paolo Di Vecchia, Lance Dixon, Riccardo Gonzo, David Nichols, Donal O'Connell, Matteo Sergola and Fei Teng for discussions. 
JPM thanks Stefano De Angelis and Vincent He for collaboration on related topics.
AG is supported by a Royal Society funding, URF\textbackslash{R}\textbackslash221015. VG is supported by Fundacao para a Ciencia e a Tecnologia (FCT) under the grant CEECIND/03356/2022, by FCT grant 2024.00230.CERN and HORIZON-MSCA-2023-SE-01-101182937-HeI. Centro de Fisica do Porto is partially funded by Fundacao para a Ciencia e a Tecnologia (FCT) under the grant UID04650-FCUP. 
CH thanks the Galileo Galilei Institute for Theoretical Physics for the hospitality and the INFN for partial support during the completion of this work.

\newpage 

\pagebreak
\widetext
\begin{center}
\textbf{\large Supplemental Material}
\end{center}

\section{Spin-Weighted Spherical Harmonics}
\label{app:SWSH}

In this appendix, we collect our conventions for spherical harmonics and recall some useful properties \cite{Blanchet:1985sp,Blanchet:1986dk,Blanchet:2013haa}.
We start by constructing a basis of symmetric trace-free (STF) tensors using the reference spatial vectors in Eq.~\eqref{eq:txyrefs} letting $\eta^{\pm} = \hat{x} \pm i \hat{y}$ and $\zeta = \hat{z}$.
The basis elements are defined as follows, for $0\le m \le \ell$, 
\begin{equation}
	Y^{\ell m}_{i_1\cdots i_\ell} = \eta^{+}_{\langle i_1}\cdots \eta^{+}_{i_m} \zeta_{i_{m+1}}\cdots \zeta_{i_\ell\rangle}\,,\qquad
	Y^{\ell (-m)}_{i_1\cdots i_\ell} = \eta^{-}_{\langle i_1}\cdots \eta^{-}_{i_m} \zeta_{i_{m+1}}\cdots \zeta_{i_\ell\rangle}\,,
\end{equation}
where angular brackets denote the STF projection. In the frame defined by the time direction $t^\mu=(1,0,0,0)$, we let
\begin{equation}
	n^\mu = (1, \sin\theta\, \cos\phi, \sin\theta \,\sin\phi,\cos\theta)
\end{equation}
as well as
\begin{equation}
	\varepsilon^{\mu}_\theta = \frac{\partial n^\mu}{\partial \theta}\,,\qquad 
	\varepsilon^{\mu}_\phi = \frac{1}{\sin\theta}\,\frac{\partial n^\mu}{\partial \phi}\,,\qquad
	\varepsilon_{\pm}^{\mu} = \frac{1}{\sqrt2}\left(
	\varepsilon_\theta^\mu \pm i \varepsilon_\phi^{\mu}
	\right).
\end{equation}

Spin-$s$ spherical harmonics are defined for integer $s\ge 0$ by
\begin{equation}\label{eq:SWSH}
Y^{\ell m}_{\pm s} = \mathcal{N}_{s}^{\ell m} \varepsilon_{\pm}^{i_1}\cdots \varepsilon_{\pm}^{i_s}\, n^{i_{s+1}}\cdots n^{i_\ell} Y_{i_1\cdots i_\ell}^{\ell m}\,.
\end{equation}
The normalization factors read as follows,
\begin{equation}\label{eq:calNs}
	\mathcal{N}_s^{\ell m} = \ell!(2\ell + 1)!!\,  \sqrt{\frac{2^{s}}{4\pi (2\ell+1)(\ell+m)! (\ell-m)! (\ell-s)!(\ell+s)!}}\,(-1)^{\theta(m)m} \qquad
	\text{for }s=0,1,2\,.
\end{equation}
The spin-weighted spherical harmonics \eqref{eq:SWSH} 
obey the orthonormality condition
\begin{equation}
	\oint Y_{\pm s}^{\ell m \ast}\,Y_{\pm s}^{\ell^\prime m^\prime} d\Omega= \delta^{\ell \ell^\prime} \delta^{m m^\prime}\,,
\end{equation}
where $d\Omega = \sin\theta\,d\theta\,d\phi$, and allow one to decompose the two physical polarizations of the waveform according to
\begin{equation}
w_{\pm}^{\ell m} = \oint Y^{\ell m \ast }_{\pm 2} \varepsilon^{\mu}_{\pm} w_{\mu\nu} \varepsilon^{\nu}_{\pm} \,d\Omega
\,,\qquad
   \varepsilon^{\mu}_{\pm} w_{\mu\nu} \varepsilon^{\nu}_{\pm}=
\sum_{\ell=2}^\infty 
\sum_{m=-\ell}^\ell
w^{\ell m}_\pm \,Y^{\ell m}_{\pm 2}\,.
\end{equation}

\section{Multipole expansion of the linear memory and logarithmic corrections}
\label{app:multipolelinear}

The multipole projection of any of the building blocks $p_a^\mu p_a^\nu/p_a\cdot n$ of the linear memory can be easily written down in a generic frame for any $\ell$, $m$.
This follows from Eq.~(100) of Ref.~\cite{Fucito:2024wlg} as we now describe. For any future-directed timelike vector $p^\mu$ such that $-p^2=\mu^2>0$ lying in the $xy$ plane, we define
\begin{equation}
    \vec{p}_t^{\ \mu}
	=
	p^\mu + t^\mu (t\cdot p)\,,
    \qquad
    \sigma_t = - \frac{t\cdot p}{\mu} = \frac{1}{\sqrt{1-v_t^2}}\,,
    \qquad
\frac{\vec{p}_t}{|\vec{p}_t |} 
    = 
    \hat{x}\,\sin\Phi_p
    +
    \hat{y}\,\cos\Phi_p\,.
\end{equation}
Then the multipoles
\begin{equation}\label{eq:linmemmultipoles}
	p_{\pm}^{\ell m} = \oint Y_{\pm 2}^{\ell m \ast}  \, \frac{(p\cdot \varepsilon_{\pm})^2}{p\cdot n} \, d\Omega
\end{equation}
are given by
\begin{equation}\label{eq:alllinmemmultipoles}
    p_{\pm}^{\ell m}
	=
	-\mu (-i)^m \pi \sigma_t v_t^\ell \ell! a^0_{\ell m} \,{}_2 F_{1}\left(\tfrac{\ell-1}{2},\tfrac{\ell}{2},\ell+\tfrac{3}{2},v_t^{2}\right)
	e^{i m\Phi_p}
\end{equation}
with 
\begin{equation}\label{eq:a0ellm}
a^0_{\ell m} 
=
\frac{2^{\ell+m} \Gamma (\ell-1) \sqrt{\frac{(\ell-1) \ell (\ell+1) (\ell+2) (2 \ell+1) (\ell-m)!}{(\ell+m)!}}}{\Gamma (2 \ell+2) \Gamma \left(\frac{1-\ell-m}{2}\right) \Gamma \left(\frac{2+\ell-m}{2} \right)}\,.
\end{equation}
Eq.~\eqref{eq:alllinmemmultipoles} computes the multipoles for each term entering the nonlinear memory $f^{\mu\nu}$ in \eqref{eq:WeinbergF} in the nonspinning case, by choosing $xy$ as the scattering plane. By expanding the result for small deflections, one then trivially obtains the multipole decompositions of the various tensor structures appearing in the PM expansion \eqref{eq:S1S3T3}.

The results of Refs.~\cite{Saha:2019tub,Sahoo:2018lxl,Sahoo:2021ctw} for $\log\omega$, $\omega(\log\omega)^2$ and the conjecture \cite{Alessio:2024onn} for $\omega^{n-1}(\log\omega)^{n}$ can be taken into account in a compact way up to and including $\mathcal{O}(G^3)$ by replacing the $f^{\mu\nu}$ in \eqref{eq:FdeltaF} with $f_\text{LL}^{\mu\nu}$ given by
\begin{equation}\label{eq:allLL}
\varepsilon_{\mu}\, f^{\mu\nu}_\text{LL}\, \varepsilon_{\nu}
=
\omega^{2iG E\omega}\,\Bigg[ 
\sum_{a=1,2}
\frac{(p_a\cdot \varepsilon_\text{CM})^2}{p_a\cdot n}\,
\omega^{i G E h(\sigma) \omega}
+
\sum_{a=3,4}
\frac{(p_a\cdot \varepsilon_\text{CM})^2}{p_a\cdot n}\,
\omega^{-i G E h(\sigma) \omega}
\Bigg],
\end{equation}
where $E = (p_1+p_2)\cdot n$ and $h(\sigma) = \frac{\sigma(2\sigma^2-3)}{(\sigma^2-1)^{3/2}} $,
while $\varepsilon_\text{CM}^\alpha = \varepsilon^\alpha - n^\alpha\,\varepsilon\cdot (p_1+p_2)/E$.
In the \emph{center-of-mass} frame, all exponential factors in \eqref{eq:allLL} are angle-independent and $\varepsilon\cdot (p_1+p_2)=0$, so that the multipole expansion for any leading log is immediately obtained via \eqref{eq:linmemmultipoles} from \eqref{eq:alllinmemmultipoles} for generic $\ell$, $m$.
This also coincides with the result in the rest frame of particle $1$ to leading order for small mass-ratio $m_2/m_1\ll 1$.

The multipole decomposition of \eqref{eq:allLL} can be performed as follows in a generic frame where the scattering orbits lie in the $xy$ plane. Compared to \eqref{eq:linmemmultipoles}, to each order in $\omega^{n-1}(\log\omega)^n$, projecting Eq.~\eqref{eq:allLL} onto $Y_{-2}^{\ell m}$ involves additional factors of the type $(p'\cdot n)(p''\cdot n)\cdots$, which can be written as sums of terms of the type $(\sin\theta)^{j} e^{\pm i k \phi}$, with $0\le k \le j \le n$ and $j$, $k$ both odd or both even. Then, it is convenient to decompose this as follows (here $Y^{\ell m}_{-2}=0$ for $|m|>\ell$)
\begin{equation}\label{eq:decomp}
    Y_{-2}^{\ell m}(\theta,\phi)^\ast \, (\sin\theta)^{j} e^{\pm i k \phi}
    =
    \sum_{\ell'=\ell-j}^{\ell+j} c^{(j,\pm k)}_{\ell \ell' m} \, Y_{-2}^{\ell' (m\mp k)}(\theta,\phi)^\ast\,,
\end{equation}
where the coefficients $c^{(j,\pm k)}_{\ell \ell' m}$ are determined by 
\begin{equation}\label{eq:findcjkllpm}
c^{(j,\pm k)}_{\ell \ell' m} 
=
\oint
    Y_{-2}^{\ell m}(\theta,\phi)^\ast \, (\sin\theta)^{j} e^{\pm i k \phi} \, Y_{-2}^{\ell' (m\mp k)}(\theta,\phi) \, d\Omega\,.
\end{equation}
Noting that
\begin{equation}
    (\sin \theta)^j e^{\pm i k \phi} =  
    \frac{Y_0^{\ell_\pm\ell_\pm}(\theta,\phi)
    Y_0^{\ell_\mp(-\ell_\mp)}(\theta,\phi)}{Y_0^{\ell_\pm\ell_\pm}(\tfrac{\pi}{2},0)
    Y_0^{\ell_\mp(-\ell_\mp)}(\tfrac{\pi}{2},0)} \quad \text{with} \quad \ell_\pm=\frac{j\pm k}{2}\,,
\end{equation}
we can also perform the decomposition
\begin{equation}\label{eq:decompjkasY}
    (\sin \theta)^j e^{\pm i k \phi} = \sum_{q=k}^j d^{(j,\pm k)}_q Y^{q (\pm k)}_0(\theta,\phi)
\end{equation}
with
\begin{equation}\label{eq:finddjkq}
    d^{(j,\pm k)}_q = \frac{1}{Y_0^{\ell_\pm\ell_\pm}(\tfrac{\pi}{2},0)
    Y_0^{\ell_\mp(-\ell_\mp)}(\tfrac{\pi}{2},0)}\oint Y_0^{\ell_\pm\ell_\pm}(\theta,\phi)
    Y_0^{\ell_\mp(-\ell_\mp)}(\theta,\phi)Y^{q (\pm k)}_0(\theta,\phi)^\ast 
     \,d\Omega\,.
\end{equation}
Substituting \eqref{eq:decompjkasY} into \eqref{eq:findcjkllpm}, we obtain
\begin{equation}\label{eq:findcjkllpmEXPLICIT}
    c^{(j,\pm k)}_{\ell \ell' m} 
=
\sum_{q=k}^j
d^{(j,\pm k)}_q
\oint
    Y_{-2}^{\ell m}(\theta,\phi)^\ast \, 
    Y_0^{q (\pm k)}(\theta,\phi)\,
    Y_{-2}^{\ell' (m\mp k)}(\theta,\phi) \, d\Omega
\end{equation}
The ``triple integrals'' appearing in Eqs.~\eqref{eq:finddjkq} and \eqref{eq:findcjkllpmEXPLICIT} are the spin-weighted Clebsch--Gordan coefficients,
\begin{equation}
\oint
    Y_{s_1}^{\ell_1 m_1}(\theta,\phi) \, Y_{s_2}^{\ell_2 m_2}(\theta,\phi)\,
    Y_{s_3}^{\ell_3 m_3}(\theta,\phi) \, d\Omega
    =
    C^{\ell_1 \ell_2 \ell_3, m_1 m_2 m_3}_{s_1 s_2 s_3}\,,
\end{equation}
where we recall that $Y^{\ell m}_s(\theta,\phi)=(-1)^m Y^{\ell(-m)}_{-s}(\theta,\phi)^\ast$ and, introducing Wigner's 3-$j$ symbol,
\begin{equation}
C^{\ell_1 \ell_2 \ell_3, m_1 m_2 m_3}_{s_1 s_2 s_3}=\sqrt{\frac{(2\ell_1+1)(2\ell_2+1)(2\ell_3+1)}{4\pi}}
\left(
\begin{array}{ccc}
    \ell_1 & \ell_2 & \ell_3  \\
    m_1 & m_2 & m_3
\end{array}
\right)
\left(
\begin{array}{ccc}
    \ell_1 & \ell_2 & \ell_3 \\
    -s_1 & -s_2 & -s_3
\end{array}
\right)
\end{equation}
for $s_1+ s_2+s_3=0$.
So, 
\begin{equation}
    d_q^{(j,\pm k)}=
    \frac{(-1)^k C^{\ell_{\pm}\ell_{\mp}q,\ell_{\pm}(-\ell_{\mp})(\mp k)}_{000}}{Y_0^{\ell_\pm\ell_\pm}(\tfrac{\pi}{2},0)
    Y_0^{\ell_\mp(-\ell_\mp)}(\tfrac{\pi}{2},0)}
\end{equation}
and
\begin{equation}
    c^{(j,\pm k)}_{\ell \ell' m}=
    \sum_{q=k}^j
    \frac{(-1)^k C^{\ell_{\pm}\ell_{\mp}q,\ell_{\pm}(-\ell_{\mp})(\mp k)}_{000} (-1)^m C^{\ell q \ell',(-m) (\pm k) (m\mp k)}_{2 0 (-2)}}{Y_0^{\ell_\pm\ell_\pm}(\tfrac{\pi}{2},0)
    Y_0^{\ell_\mp(-\ell_\mp)}(\tfrac{\pi}{2},0)}\,.
\end{equation}
In this way, \eqref{eq:decomp} reduces the multipole decomposition of any term entering \eqref{eq:allLL} to the form \eqref{eq:linmemmultipoles} and thus to \eqref{eq:alllinmemmultipoles}.

\section{Polynomials for \texorpdfstring{$\ell\ge3$}{ell>=3}}
\label{app:polynomials}

Here we provide the explicit expressions for the polynomials entering the nonlinear memory multipoles \eqref{eq:deltaFlmexplicit} as in \eqref{eq:calFdefinition} for $\ell\ge3$. All results are also collected in a computer-friendly format in the ancillary file \texttt{nlmMultipoles.m}, which consists of a list of lists whose entry $(\ell-1, \ell+1+m)$ gives the explicit expression for $\delta F^{\ell m} (G^3 \pi^2 m_1^2 m_2^2 b^{-3} \mathcal{N}_2^{\ell m})^{-1}$ for $\ell=2,\ldots,5$ and $m=-\ell,-\ell+1,\ldots,\ell-1,\ell$. 
\begin{subequations}
	\begin{align}
		\begin{split}
			f_1^{33} &= \frac{1}{11520}(-16800 \sigma ^9+911914 \sigma ^8-1263600 \sigma ^7-1816089 \sigma ^6+1459568 \sigma ^5+4851356 \sigma ^4
            \\
            &-6058128 \sigma ^3+3151511 \sigma ^2-1977680 \sigma +806248)
		\end{split}
		\\ 
		f_2^{33} &= \frac{1}{24}(\sigma^2-1)(35 \sigma ^7+1154 \sigma ^6-1155 \sigma ^5-1068 \sigma ^4+2605 \sigma ^3+402 \sigma ^2+147 \sigma -1672)\\
		\begin{split}
		f_3^{33} &=-\frac{\sigma}{768}(1120 \sigma ^9+35008 \sigma ^8-23280 \sigma ^7-115040 \sigma ^6+56368 \sigma ^5+135027 \sigma ^4-40464 \sigma ^3-64654 \sigma ^2
        \\
        &+6768 \sigma +12367)
		\end{split}
	\end{align}
\end{subequations}

\begin{subequations}
	\begin{align}
		\begin{split}
			f_1^{31} &= \frac{1}{11520} (-3360 \sigma ^9-233722 \sigma ^8+253584 \sigma ^7+761313 \sigma ^6-893264 \sigma ^5-691076 \sigma ^4+605040 \sigma ^3
            \\
            &+332737 \sigma ^2+30320 \sigma -172072)
		\end{split}
		\\ 
		f_2^{31} &=\frac{1}{120}(\sigma^2-1)(35 \sigma ^7-2020 \sigma ^6+705 \sigma ^5+3666 \sigma ^4-7931 \sigma ^3+1224 \sigma ^2-969 \sigma +3050)\\
		\begin{split}
		f_3^{31} &=-\frac{\sigma}{3840}(1120 \sigma ^9-66560 \sigma ^8+36240 \sigma ^7+188800 \sigma ^6-99728 \sigma ^5-239367 \sigma ^4+73200 \sigma ^3
        \\
        &+147950 \sigma ^2-13392 \sigma -31763)
		\end{split}
	\end{align}
\end{subequations}

\begin{subequations}
	\begin{align}
		\begin{split}
			f_1^{44} &= \frac{1}{161280}(-117600 \sigma ^{10}+41618132 \sigma ^9-37975440 \sigma ^8-109303456 \sigma ^7+61852864 \sigma ^6+195002541 \sigma ^5
            \\
            &-138084896 \sigma ^4-78767544 \sigma ^3+15535392 \sigma ^2+109431017 \sigma -58404560)
		\end{split}
		\\ 
		f_2^{44} &= \frac{1}{48}(\sigma^2-1)(35 \sigma ^8+8636 \sigma ^7-2040 \sigma ^6-15156 \sigma ^5+5766 \sigma ^4+24116 \sigma ^3+7056 \sigma ^2-23612 \sigma -449) \\
		\begin{split}
		f_3^{44} &= \frac{1}{1536}(-1120 \sigma ^{11}-274432 \sigma ^{10}+43920 \sigma ^9+943232 \sigma ^8-69184 \sigma ^7-1491926 \sigma ^6-34272 \sigma ^5+1619841 \sigma ^4
        \\
        &+76320 \sigma ^3-1146862 \sigma ^2-17712 \sigma +344705)
		\end{split}
	\end{align}
\end{subequations}

\begin{subequations}
	\begin{align}
		\begin{split}
			f_1^{42} &= \frac{1}{161280}(-16800 \sigma ^{10}-30914056 \sigma ^9+22702800 \sigma ^8+86747858 \sigma ^7-54426272 \sigma ^6-116220243 \sigma ^5\\
            &
            +57735808 \sigma ^4+114882867 \sigma ^3-65687616 \sigma ^2-54553756 \sigma +39584560)
		\end{split}
		\\ 
		f_2^{42} &=\frac{1}{336}(\sigma^2-1)(35 \sigma ^8-46816 \sigma ^7+750 \sigma ^6+91068 \sigma ^5-5976 \sigma ^4-110680 \sigma ^3-34566 \sigma ^2+87484 \sigma +3469) \\
		\begin{split}
		f_3^{42} &= \frac{1}{10752}(-1120 \sigma ^{11}+1500032 \sigma ^{10}-45360 \sigma ^9-5117632 \sigma ^8-127840 \sigma ^7+8221576 \sigma ^6+603456 \sigma ^5
        \\
        &-8688171 \sigma ^4-532416 \sigma ^3+5827397 \sigma ^2+110448 \sigma -1739380)
		\end{split}
	\end{align}
\end{subequations}

\begin{subequations}
	\begin{align}
		\begin{split}
			f_1^{40} &= \frac{1}{107520}(-6720 \sigma ^{10}+18720268 \sigma ^9-13137248 \sigma ^8-51452252 \sigma ^7+31032448 \sigma ^6+66842569 \sigma ^5
            \\
            &-35896768 \sigma ^4-62973174 \sigma ^3+41068224 \sigma ^2+28885689 \sigma -23059936)
		\end{split}
		\\ 
		f_2^{40} &= \frac{1}{560}(\sigma^2-1)^2(35 \sigma ^6+69960 \sigma ^5+1715 \sigma ^4-67784 \sigma ^3-31335 \sigma ^2+95744 \sigma +6545)\\
		\begin{split}
		f_3^{40} &= \frac{1}{35840}(-2240 \sigma ^{11}-4473600 \sigma ^{10}-150240 \sigma ^9+15619200 \sigma ^8+1187456 \sigma ^7-24641350 \sigma ^6
        \\
        &-2470336 \sigma ^5+25244275 \sigma ^4+1785664 \sigma ^3-16815540 \sigma ^2-350304 \sigma +5059315)
		\end{split}
	\end{align}
\end{subequations}

\begin{subequations}
	\begin{align}
		\begin{split}
			f_1^{55} &= \frac{1}{17203200}(-7526400 \sigma ^{11}+10702323340 \sigma ^{10}-8533785600 \sigma ^9-28927624672 \sigma ^8+15410782208 \sigma ^7
            \\
            &+81973607807 \sigma ^6-52988807168 \sigma ^5-111199134150 \sigma ^4+66405210112 \sigma ^3+79819266827 \sigma ^2
            \\
            &-47769705472 \sigma -4789880032)
		\end{split}
		\\ 
        \begin{split}
		f_2^{55} &= \frac{1}{160}(\sigma^2-1)(70 \sigma ^9+71035 \sigma ^8-6400 \sigma ^7-132076 \sigma ^6+17900 \sigma ^5+460770 \sigma ^4+80080 \sigma ^3
        \\
        &-507500 \sigma ^2+2430 \sigma +67451)
        \end{split}
        \\
		\begin{split}
		f_3^{55} &= - \frac{\sigma}{163840}(71680 \sigma ^{11}+72616960 \sigma ^{10}-4572160 \sigma ^9-248627712 \sigma ^8-1605632 \sigma ^7+653460698 \sigma ^6
        \\
        &+28141568 \sigma ^5-1181797807 \sigma ^4-27310080 \sigma ^3+1026851480 \sigma ^2+5667840 \sigma -321994675)
		\end{split}
	\end{align}
\end{subequations}

\begin{subequations}
	\begin{align}
		\begin{split}
			f_1^{53} &= \frac{1}{154828800}(-7526400 \sigma ^{11}-51717468452 \sigma ^{10}+37409879040 \sigma ^9+142504354880 \sigma ^8-87045062656 \sigma ^7
            \\
            &-343992349165 \sigma ^6+208672307200 \sigma ^5+532280290818 \sigma ^4-349860841472 \sigma ^3-305427849601 \sigma ^2
            \\
            &+190707381248 \sigma +26268908960)
		\end{split}
		\\ 
        \begin{split}
		f_2^{53} &=\frac{1}{1440}(\sigma^2-1)(70 \sigma ^9-347873 \sigma ^8+1040 \sigma ^7+667604 \sigma ^6+18428 \sigma ^5-1880886 \sigma ^4
        \\
        &-304256 \sigma ^3+2032372 \sigma ^2+2478 \sigma -350257) 
        \end{split}\\
		\begin{split}
		f_3^{53} &= -\frac{\sigma}{1474560}(71680 \sigma ^{11}-356344832 \sigma ^{10}+3046400 \sigma ^9+1213687296 \sigma ^8+53370880 \sigma ^7-3009375502 \sigma ^6
        \\
        &-164214784 \sigma ^5+5167571357 \sigma ^4+133687296 \sigma ^3-4409331256 \sigma ^2-27141120 \sigma +1392991865)
		\end{split}
	\end{align}
\end{subequations}

\begin{subequations}
	\begin{align}
		\begin{split}
			f_1^{51} &= \frac{1}{103219200}(-2150400 \sigma ^{11}+10930945292 \sigma ^{10}-7724544000 \sigma ^9-29725155824 \sigma ^8
            \\
            &+18642030592 \sigma ^7+66548795719 \sigma ^6
            -42886721536 \sigma ^5-100387851678 \sigma ^4+68339449856 \sigma ^3
            \\
            &+58341360475 \sigma ^2-36368064512 \sigma -5689042784)
		\end{split}
		\\ 
        \begin{split}
		f_2^{51} &= \frac{1}{1680}(\sigma^2-1)^2(35 \sigma ^7
        +127859 \sigma ^6+2415 \sigma ^5-120129 \sigma ^4-57719 \sigma ^3+498249 \sigma ^2+8229 \sigma -134587)
        \end{split}\\
		\begin{split}
		f_3^{51} &= -\frac{\sigma}{6881280}(143360 \sigma ^{11}+523464704 \sigma ^{10}+13711360 \sigma ^9-1809868800 \sigma ^8-122871808 \sigma ^7+4213651606 \sigma ^6
        \\
        &+266874880 \sigma ^5-6976891697 \sigma ^4-196841472 \sigma ^3+5947322992 \sigma ^2+38983680 \sigma -1896408725)
		\end{split}
	\end{align}
\end{subequations}

\end{document}